\documentclass[twocolumn]{revtex4}
\usepackage{amsmath,amsbsy,amssymb,graphicx,color,pifont}
\usepackage[mathscr]{euscript}

\def\babc{\begin{subequations}}
\def\eabc{\end{subequations}}
\def\be{\begin{equation}}
\def\ee{\end{equation}}
\def\ba{\begin{array}}
\def\ea{\end{array}}
\def\nn{\nonumber}
\def\hh{\hspace*{0.25mm}}
\def\h{\hspace*{0.5mm}}
\def\g{\textcolor[rgb]{0.8,0.8,0.8}}
\def\sinh{{\tt sh}}
\def\cosh{{\tt ch}}
\def\sin{{\tt sin}}
\def\cos{{\tt cos}}

\begin{document}

\title{Edge states in 2D lattices with hopping anisotropy and Chebyshev polynomials}

\author{M. Eliashvili$^{1,2}$, G.I. Japaridze$^3$, G. Tsitsishvili$^{1,2}$\footnote{giorgi.tsitsishvili@tsu.ge} and G. Tukhashvili$^1$}
\affiliation{$^1$Faculty of Exact and Natural Sciences, Tbilisi State University Chavchavadze Ave. 3, Tbilisi 0179 Georgia\\
$^2$Razmadze Mathematical Institute, Tbilisi State University, Tamarashvili Str. 6, Tbilisi 0177 Georgia \\
$^3$College of Engineering, Ilia State University, Cholokashvili Ave. 3-5, Tbilisi 0162 Georgia}

\begin{abstract}
Analytic technique based on Chebyshev polynomials is developed for studying two-dimensional lattice ribbons with hopping anisotropy.
In particular, the tight-binding models on square and triangle lattice ribbons are investigated with anisotropic nearest neighbouring hoppings.
For special values of hopping parameters the square lattice becomes topologically equivalent to a honeycomb one either with zigzag or armchair
edges. In those cases as well as for triangle lattices we perform the exact analytic diagonalization of tight-binding Hamiltonians in terms of
Chebyshev polynomials. Deep inside the edge state subband the wave functions exhibit exponential spatial damping which turns into power-law
damping at edge-bulk transition point. It is shown that strong hopping anisotropy crashes down edge states, and the corresponding critical conditions
are found.
\end{abstract}

\maketitle

\section{Introduction}

The concept of edge states dates back to Tamm \cite{tamm} who pointed out in 1932 that the energy levels of a crystal can give birth to "surface states"
where electrons are localized along the crystal surface. Subsequent studies of the issue were carried out by different authors \cite{rij,maue,goodwin,shock}
till late 1930's.

The physics of edge states acquired new life in last decades due to the progress in fabrication of low-dimensional electron structures and novel materials.
Current carrying edge states play decisive role in the formation of integer \cite{qhe_1a,qhe_1b,streda} and fractional \cite{qhe_2a,qhe_2b,qhe_2c}
quantum Hall states observed in GaAs heterostructures, oxides heterostructures \cite{qhe-in-Oxides_1,qhe-in-Oxides_2,qhe-in-Oxides_3} and in
graphene \cite{qhe-in-graphene_1,qhe-in-graphene_2,qhe-in-graphene_3}. Interest in physics of edge states has been considerably heated up by the discovery
of topological insulators \cite{TI_1,TI_2}. These are systems with insulating bulk and topologically protected conducting edge states (see Ref. [20] for recent review).
One can exemplify other physical systems e.g. optical lattices \cite{opt-lat} and photonic crystals \cite{phot-crys} where the edge states do emerge.

Edge states were usually studied in 2D lattice electron systems and within the framework of tight-binding models \cite{schweizer,macdonald,streda,sun},
though the Dirac equation approaches have been also carried out \cite{fertig,Moradinasab_2012} (see Ref. [28-30] for more mathematical treatment).

After seminal theoretical papers by Fujita et al. \cite{Fujita_1996a} it became clear that edges have strong impact on the low-energy electronic
structure and electronic transport properties of nanometer-sized graphene ribbons \cite{Fujita_1996a,Fujita_1996b,Fujita_1998,Fujita_1999}.
Because edge states substantially determine infrared transport and magnetic properties of graphene nanoribbons, considerable efforts were devoted
during the last decade to studying the effect of edges in graphitic nanomaterials (see Ref. [35] for review).

Synthesis of two-dimensional boron nanoribbons with triangular crystal structure has been reported recently \cite{Boron_Ribbon_1}.
Theoretical estimates show that monolayers of a boron built up of triangular and hexagonal structural elements are energetically more stable than
the flat triangular sheets \cite{Boron_Ribbon_2}. Therefore general perception of a monolayer boron sheet is that it occurs as a buckled sheet with
triangular and hexagonal components. As a result electronic band structure of boron nanoribbons with mixed structure has become the subject of subsequent
theoretical and numerical analysis \cite{Boron_Ribbon_3} while the edge states in pure triangular ribbons have not been studied in details.

In this paper we consider tight-binding models of free electrons living on two-dimensional square and triangular lattice ribbons. In the case of square-lattice
ribbon electron delocalization process is characterized by four different hopping parameters $t_u$, $t_d$, $t_l$, $t_r$, while in the case of triangular-lattice
ribbon -- by three different hopping parameters $t_1$, $t_2$, $t_3$ parameterising hoppings along the three linear directions on the triangular lattice.

In Section 2 we study the square-lattice ribbon. For the particular regimes of hopping parameters the Hamiltonian under consideration is reduced to that of an electron on
a honeycomb ribbon with either zigzag or armchair edges. For these physically important sets of hopping amplitudes we solve the eigenvalue problem exactly and express
the solutions in terms of Chebyshev polynomials. In the case of zigzag boundaries we reproduce the flat band of edge states \cite{Fujita_1996a,Fujita_1996b}.
Inclusion of hopping anisotropy allows to trace out the corresponding response of the system. In particular, we show that the formation of edge states depends on
strength of anisotropy and may not occur at all if the anisotropy between certain directions is sufficiently strong.

In Section 3 we deal with triangle-lattice ribbons. We consider three different options for edge configurations and solve the diagonalization problems in terms
of Chebyshev polynomials. Prior attention is paid to the occurrence of edge states and the corresponding necessary conditions on hopping parameters are found.

Results are summarized in Section 4. Calculational details are collected in Appendix.

\section{Anisotropic square ribbon}

In this Section we consider electrons on a square lattice shown in Fig. 1 with four different hopping amplitudes $t_u$, $t_d$, $t_l$, $t_r$.
The lattice is finite in $x$-direction comprising of $N$ one-dimensional chains, and infinite in $y$-direction.
In response to the particular hopping anisotropy the lattice is considered as consisting of two Bravais sublattices labeled by $\mu=\bullet,\circ$.
Integers $1\leqslant n\leqslant N$ and $-\infty<m<+\infty$ parameterize the unit cell indicated by dashed area in Fig. 1.
\begin{figure}[h]
\begin{center}
\includegraphics{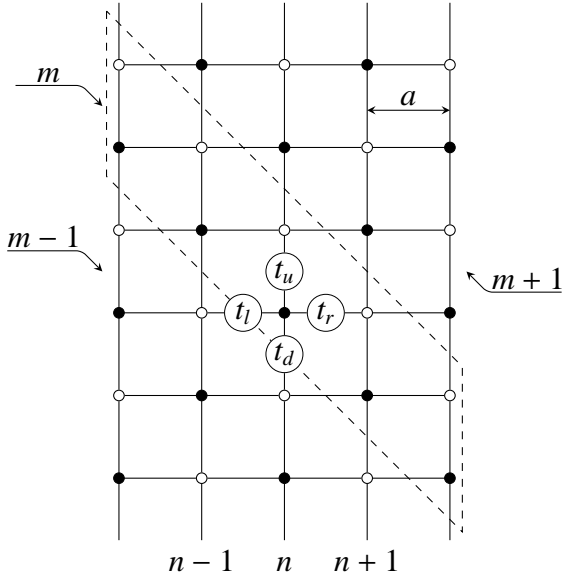}
\end{center}
\caption{Square-lattice ribbon with hopping anisotropy. Dashed area represents the unit cell. The ribbon is $y$-periodic with periodicity $2a$.}
\end{figure}

The tight-binding Hamiltonian appears as
\begin{align}
H&=t_u\sum_m\sum_{n=1}^N\Big[c^\dag_\circ(n,m)c^{}_\bullet(n,m)+h.c.\Big]+\nn\\
&+t_d\sum_m\sum_{n=1}^N\Big[c^\dag_\circ(n,m-1)c^{}_\bullet(n,m)+h.c.\Big]+\nn\\
&+t_r\sum_m\sum_{n=1}^{N-1}\Big[c^\dag_\circ(n+1,m)c^{}_\bullet(n,m)+h.c.\Big]+\nn\\
&+t_l\sum_m\sum_{n=2}^{N}\Big[c^\dag_\circ(n-1,m-1)c^{}_\bullet(n,m)+h.c.\Big]
\end{align}
where $c_\mu^\dag(n,m)$ and $c_\mu^{}(n,m)$ are electron creation and annihilation operators.

Note that the terms with $n=N$ and $n=1$ are absent in third and fourth terms of (1) respectively.
This reflects the absence of hoppings away beyond the boundaries.

Separation between the nearest sites is $a$, and the lattice is periodic in $y$-direction with the period $2a$,
hence we employ the Fourier transform in $y$-direction
\be
c^{}_\mu(n,m)=\frac{1}{\sqrt{\pi/a}}\int_{BZ}e^{+ik(2a)m}c^{}_{\mu,n}(k)dk
\ee
where the length of the Brillouin zone is $2\pi/(2a)=\pi/a$.

Introducing $\psi_\mu=(c_{\mu1},\ldots,c_{\mu N})$ and $\Psi=(\psi_\circ,\psi_\bullet)$ we rewrite the Hamiltonian (1) as
\be
H=\int\Psi^\dag(k)\hh\mathscr H(k)\hh\Psi(k)\hh dk
\ee
\be
\mathscr H=\left\lgroup\ba{cl}0 & T\\\vspace*{-1mm}\\T^\dag & 0\ea\right\rgroup
\ee
where
\be
T=t_u+t_de^{+2ika}+t_r\hh\beta^\dag+t_l\hh e^{+2ika}\beta.
\ee
Here $\beta$ is the $N\times N$ matrix
\be
\beta=\left\lgroup\ba{cccccc}
\g{0} & {\bf 1} & \g{0} & \cdots & \g{0} & \g{0}\\
\g{0} & \g{0} & {\bf 1} & \cdots & \g{0} & \g{0}\\
\g{0} & \g{0} & \g{0} & \cdots & \g{0} & \g{0} \vspace*{-1mm}\\
\vdots & \vdots & \vdots &  & \vdots & \vdots\\
\vspace*{-3.5mm}\\
\g{0} & \g{0} & \g{0} & \cdots & \g{0} & {\bf 1}\\
\g{0} & \g{0} & \g{0} & \cdots & \g{0} & \g{0}
\ea\right\rgroup.
\ee

The eigenvalue equation for $\mathscr H$ leads to the system of entangled equation
\babc
\begin{align}
(t_u+t_de^{+2ika}+t_r\hh\beta^\dag+t_l\hh e^{+2ika}\beta)\psi_\bullet&=E\psi_\circ\hh,\\
\nn\\
(t_u+t_de^{-2ika}+t_r\hh\beta+t_l\hh e^{-2ika}\beta^\dag)\psi_\circ&=E\psi_\bullet\hh.
\end{align}
\eabc

We consider three cases when this entanglement becomes soluble.

\ding{182} Instead of $\mathscr H\Psi=E\Psi$ one may consider $\mathscr H^2\Psi=E^2\Psi$
where the entanglement is absent. However, the linear combination of $\beta$ and $\beta^\dag$
involved in (7) is a tri-diagonal matrix. Consequently, the matrices appearing in $\mathscr H^2$ are
penta-diagonal and lead to five-term recurrence relations for the components of $\psi_\bullet$ and
$\psi_\circ$. Taking $t_l=0$ the penta-diagonal form of $\mathscr H^2$ turns into tri-diagonal one
and the equation $\mathscr H^2\Psi=E^2\Psi$ gives out three-term recurrence relation which appears
soluble in terms of Chebyshev polynomials. Switching off the $t_l$-hoppings in Fig. 1 the lattice turns into the
one shown in the left panel of Fig. 2 which is topologically equivalent to a honeycomb ribbon with zigzag edges.
\begin{figure}[h]
\begin{center}
\includegraphics{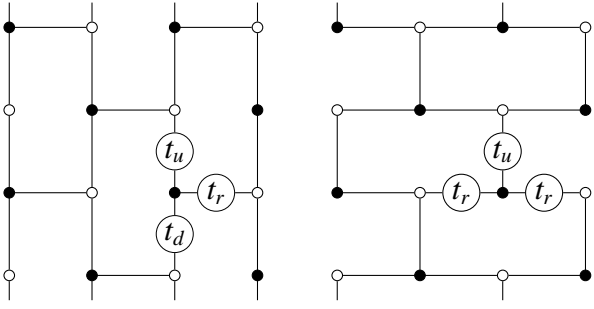}
\end{center}
\caption{Removing $t_l$-links in the initial ribbon the system turns into a honeycomb ribbon with zigzag edges (left).
Equalizing $t_l=t_r$ and putting $t_d=0$ the initial ribbon turns into an armchair edged honeycomb ribbon (right).}
\end{figure}

\ding{183} Taking $t_l=t_r$ we find $[\h T^\dag,T\h]=0$ i.e. the two matrices in the left hand sides of (7a) and (7b)
can be diagonalized simultaneously and we come to three-term recurrence relation soluble in terms of Chebyshev polynomials.
This case can be reduced further to a honeycomb with armchair edges by taking $t_d=0$ as shown in the right panel of Fig. 2.

\ding{184} We study zero modes ($E=0$) in the anisotropic square lattice. In that case the system (7) trivially decouples
into two independent equations each of three-term recurrence form.

We consider these three options separately in the following subsections.

\subsection{Zigzag honeycomb ($t_l=0$)}

For $t_l=0$ the square ribbon is topologically equivalent to a honeycomb with zigzag edges.
The eigenvalue system (7) takes the form
\babc
\begin{align}
(\xi+\beta^\dag)\psi_\bullet&=\omega\psi_\circ\hh,\\
\nn\\
(\xi^*+\beta)\psi_\circ&=\omega\psi_\bullet\hh.
\end{align}
\eabc
where $\xi=(t_u+t_de^{+2ika})/t_r$ and $\omega=E/t_r$.

Squared system appears as
\babc
\begin{align}
(\omega^2-|\xi|^2-\beta^\dag\beta-\xi\beta-\xi^*\beta^\dag)\psi_\circ&=0\\
\nn\\
(\omega^2-|\xi|^2-\beta\beta^\dag-\xi\beta-\xi^*\beta^\dag)\psi_\bullet&=0
\end{align}
\eabc
and the two equations can be solved independently.

In the matrix form these appear as
\babc
\begin{align}
\left(\ba{cccccc}
\tilde w&-\xi&\g{0}&\cdots&\g{0}&\g{0}
\\
-\xi^*&w&-\xi&\cdots&\g{0}&\g{0}
\\
\g{0}&-\xi^*&w&\cdots&\g{0}&\g{0}
\\
\vdots&\vdots&\vdots& &\vdots&\vdots
\\
\g{0}&\g{0}&\g{0}&\cdots&w&-\xi
\\
\g{0}&\g{0}&\g{0}&\cdots&-\xi^*&w
\ea\right)\psi_\circ=0\\
\nn\\
\left(\ba{cccccc}
w&-\xi&\g{0}&\cdots&\g{0}&\g{0}
\\
-\xi^*&w&-\xi&\cdots&\g{0}&\g{0}
\\
\g{0}&-\xi^*&w&\cdots&\g{0}&\g{0}
\\
\vdots&\vdots&\vdots& &\vdots&\vdots
\\
\g{0}&\g{0}&\g{0}&\cdots&w&-\xi
\\
\g{0}&\g{0}&\g{0}&\cdots&-\xi^*&\tilde w
\ea\right)\psi_\bullet=0
\end{align}
\eabc
where $\tilde w=\omega^2-|\xi|^2$ and $w=\omega^2-|\xi|^2-1$.

Secular equation determining the spectrum $\omega_1,\ldots,\omega_N$ appears as (see Appendix)
\be
U_N\bigg(\frac{w}{2|\xi|}\bigg)+\frac{1}{|\xi|}\hh U_{N-1}\bigg(\frac{w}{2|\xi|}\bigg)=0
\ee
where $U_n$ is the Chebyshev polynomials of second kind which are set by the recurrence relation
$U_n(x)=2xU_{n-1}(x)-U_{n-2}(x)$ with $U_0=1$ and $U_{-1}=0$.\cite{nist}

Since the quantities $t_u$, $t_d$, $t_r$, $k$ are all combined in $\xi$ and $\omega$,
it is reasonable to present the properties of the system in terms of these two parameters.

Fig. 3 depicts $\omega_1,\ldots,\omega_N$ versus $|\xi|$ for $N=5$ and $N=13$.

Employing the technique described in Appendix we solve (10a) and (10b) separately and obtain
\babc
\begin{align}
\psi_{\circ n}&=e^{-i(n-1)\vartheta}\bigg(U_{n-1}+\frac{1}{|\xi|}\hh U_{n-2}\bigg)\psi_{\circ1}\\
\nn\\
\psi_{\bullet n}&=e^{+i(N-n)\vartheta}\bigg(U_{N-n}+\frac{1}{|\xi|}U_{N-n-1}\bigg)\psi_{\bullet N}
\end{align}
\eabc
where $\vartheta=arg(\xi)$ and $U_n\equiv U_n(\frac{w}{2|\xi|})$.

Expressions (12) are obtained by solving the homogeneous equations (10) and therefore comprise free constants
$\psi_{\bullet1}$ and $\psi_{\circ N}$. Equations (8) interrelate them as
\be
e^{+iN\vartheta}\psi_{\circ1}+\omega\hh U_N\psi_{\bullet N}=0
\ee
and the remnant free one is fixed by normalization.

We show that the states located within the shaded area in Fig. 3 are bulk states,
and the ones left beyond are edge states.

\begin{figure}[h]
\begin{center}
\includegraphics{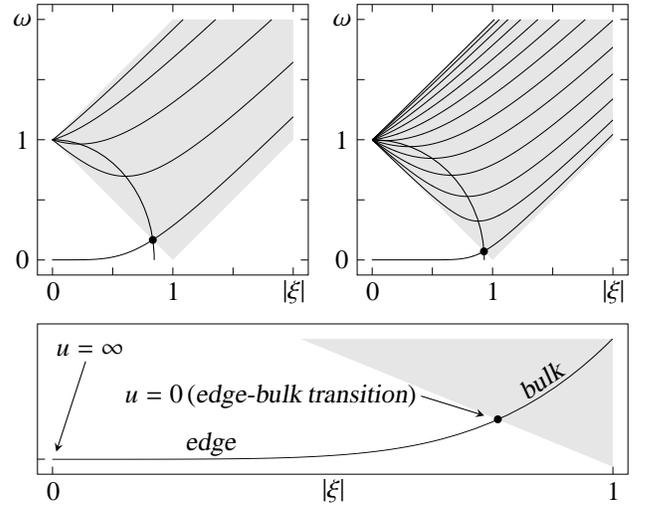}
\end{center}
\caption{Dispersion $\omega(|\xi|)$ for $N=5$ (top left) and $N=13$ (top right). Only the positive subbands are shown.
Curves across the energy bands represent the ellipse set by (18). Lower panel shows the edge subband in more details.}
\end{figure}

\subsubsection{Bulk states}

Shaded area shown in Fig. 3 is bounded from three sides by $\omega=|\xi|\pm1$ and $\omega=1-|\xi|$,
which imply that in the interior of this area we have
\be
-1\leqslant\frac{\omega^2-|\xi|^2-1}{2|\xi|}\leqslant+1.
\ee
Denoting $\frac{\omega^2-|\xi|^2-1}{2|\xi|}=\cos v$ we use the relation
\be
U_n(\cos v)=\frac{\sin[(n+1)v]}{\sin v}.
\ee
This allows to write the eigenstates (12) as
\babc
\begin{align}
\bigg|\frac{\psi_{\circ n}}{\psi_{\circ1}}\bigg|&=\frac{\sin[nv]}{\sin v}+\frac{\sin[(n-1)v]}{|\xi|\hh\sin v}\\
\nn\\
\bigg|\frac{\psi_{\bullet n}}{\psi_{\bullet N}}\bigg|&=\frac{\sin[(N-n+1)v]}{\sin v}+\frac{\sin[(N-n)v]}{|\xi|\hh\sin v}
\end{align}
\eabc
where from the oscillating behaviour with respect to $n$ is evident.
Consequently, none of the states represented by the interior of shaded area can be localized at boundaries ($n=1$ and $n=N$).
These are all bulk states.

Differentiating (11) we find
\be
\frac{d\omega}{d|\xi|}=\frac{\omega}{|\xi|}\h\frac{N\omega^2+(N+2)|\xi|^2-N}{(2N+1)\omega^2+|\xi|^2-1}
\ee
where we used $(x^2-1)U'_n(x)=nxU_n(x)-(n+1)U_{n-1}(x)$ and $U_{n+1}(x)=2xU_n(x)-U_{n-1}(x)$ together with (11).

From (17) it follows that the extrema of subbands (numerator vanishes) are located along the ellipsis set by
\be
\omega^2+\frac{N+2}{N}\h|\xi|^2=1.
\ee

Alongside with the extrema there is an extra point (indicated in bold) where the ellipsis intersects the energy bands.
As shown in the next subsection this represents the edge-bulk transition points, and the subbands located beyond the
shaded area are edge states.

\subsubsection{Edge states}

The only energy band left beyond the shaded area is the one shown in Fig. 3. In this case we have
\be
-\cosh u\stackrel\equiv\h\frac{\omega^2-|\xi|^2-1}{2|\xi|}\leqslant-1.
\ee

Taking $v=iu$ in (15) we obtain
\be
U_n(\cosh u)=\frac{\sinh[(n+1)u]}{\sinh u}.
\ee

Using (20) in secular equation (11) we find
\be
|\xi|=\frac{\sinh(Nu)}{\sinh[(N+1)u]}
\ee
which substituted into (19) leads to
\be
\omega^2=\frac{\sinh^2(Nu)}{\sinh^2[(N+1)u]}-\frac{2\hh\cosh u\h\sinh(Nu)}{\sinh[(N+1)u]}+1.
\ee
Expressions (21) and (22) set the function $\omega(|\xi|)$ parameterically via $0\leqslant u<\infty$.

Employing (20) and (21) in (12) we obtain
\babc
\begin{align}
\frac{\psi_{\circ n}}{\psi_{\circ 1}}&=e^{-i(n-1)\vartheta}\h\frac{\sinh[(N-n+1)u]}{\sinh(Nu)}\\
\nn\\
\frac{\psi_{\bullet n}}{\psi_{\bullet N}}&=e^{+i(N-n)\vartheta}\h\frac{\sinh(nu)}{\sinh(Nu)}
\end{align}
\eabc
These are depicted in Fig. 4.
\begin{figure}[h]
\begin{center}
\includegraphics{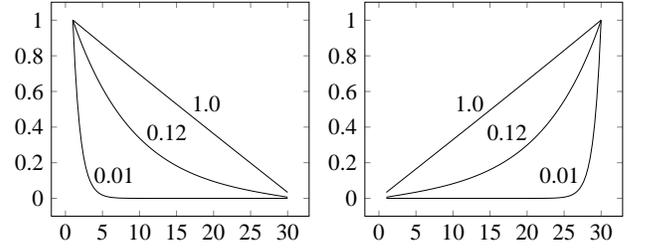}
\end{center}
\caption{Moduli of the wave functions $\big|\frac{\psi_{\circ n}}{\psi_{\circ 1}}\big|$ (left) and
$\big|\frac{\psi_{\bullet n}}{\psi_{\bullet N}}\big|$ (right) versus $n$ for $N=30$ and $u=1.0,0.12,0.01$.}
\end{figure}

From (20) and (22) we find $\omega^2U_N^2=1$. Then (13) gives
\begin{align}
|\Psi|^2&=\sum_{n=1}^N|\psi_{\circ n}|^2+\sum_{n=1}^N|\psi_{\bullet n}|^2=\nn\\
\nn\\
&=\frac{\sinh[(2N+1)u]-(2N+1)\sinh u}{2\hh\sinh u\h\sinh^2(Nu)}\h|\psi_{\circ1}|^2
\end{align}
where from we fix the value of $|\psi_{\circ1}|$ so that $|\Psi|=1$.

Taking $u=0$ in (21) and (22) we find
\babc
\begin{align}
|\xi|_{\tt cr}&=\frac{N}{N+1}\\
\nn\\
|\omega|_{\tt cr}&=\frac{1}{N+1}
\end{align}
\eabc
which represents the edge-bulk transition point indicated in bold in Fig. 3.

So far we discussed the properties with respect to $|\xi|$, while the physical variable is the momentum $k$.
Varying $k$ within the Brillouin zone the quantity $|\xi|$ varies in the interval
\be
\frac{|t_u-t_d|}{t_r}\leqslant|\xi|\leqslant\frac{t_u+t_d}{t_r}.
\ee
Therefore, occurrence of edge states depends on the values of $t_u$, $t_d$, $t_r$ as follows
\begin{itemize}
\item[$\bullet$] For $|t_u-t_d|>t_r|\xi|_{\tt cr}$ edge states never emerge.
\item[$\bullet$] For $t_u+t_d<t_r|\xi|_{\tt cr}$ edge states do emerge but never turn into bulk states.
\item[$\bullet$] For $|t_u-t_d|\leqslant t_r|\xi|_{\tt cr}\leqslant t_u+t_d$ edge states do emerge and the system exhibits the edge-bulk transition.
\end{itemize}

\subsection{Left-right isotropic case ($t_l=t_r$)}

In this case we take advantage of $[T^\dag,T]=0$, hence the two matrices can be diagonalized
simultaneously. We thus avoid the "square up" trick, i.e. are faced with three-term
recurrence relation which is soluble in terms of same polynomials.

Introduce $\psi_\bullet=e^{-\frac{i}{2}ka}G\phi_\bullet$ and $\psi_\circ=e^{+\frac{i}{2}ka}G\phi_\circ$
where the matrix $G$ is given by
\be
G=diag\big(e^{-ika},e^{-2ika},\cdots,e^{-iNka}\big)
\ee
Using $G^\dag\beta\hh G=e^{-ika}\beta$ we rewrite (7) as
\babc
\begin{align}
(t_r[\beta+\beta^\dag]+t_ue^{-ika}+t_de^{+ika})\phi_\bullet&=E\phi_\circ\hh,\\
\nn\\
(t_r[\beta+\beta^\dag]+t_ue^{+ika}+t_de^{-ika})\phi_\circ&=E\phi_\bullet\hh.
\end{align}
\eabc
i.e. we can employ the eigenstates of $\beta+\beta^\dag$. These are
\babc
\begin{align}
&(\beta+\beta^\dag)f_j=2\cos\frac{\pi j}{N+1}\h f_j\\
\nn\\
&(f_j)_n=U_{n-1}\bigg(\cos\frac{\pi j}{N+1}\bigg)=\frac{\sin\big(\frac{\pi jn}{N+1}\big)}{\sin\big(\frac{\pi j}{N+1}\big)}
\end{align}
\eabc
where $j=1,2,\ldots,N$ enumerates the eigenstates.

We put $\phi_\circ=A_\circ f_j$ and $\phi_\bullet=A_\bullet f_j$ reducing (28) to
\babc
\begin{align}
\bigg(2t_r\cos\frac{\pi j}{N+1}+t_ue^{-ika}+t_de^{+ika}\bigg)A_\bullet&=EA_\circ,\\
\nn\\
\bigg(2t_r\cos\frac{\pi j}{N+1}+t_ue^{+ika}+t_de^{-ika}\bigg)A_\circ&=EA_\bullet.
\end{align}
\eabc
Then the solubility condition leads to
\begin{align}
\frac{E_j^2}{4\hh t_r^2}
&=\frac{(t_u-t_d)^2}{4\hh t_r^2}\sin^2(ka)+\nn\\
&+\bigg[\cos\frac{\pi j}{N+1}+\frac{t_u+t_d}{2\hh t_r}\cos(ka)\bigg]^2.
\end{align}

The eigenstates (29b) oscillate with respect to $n$. Hence, in the square lattice with $t_l=t_r$
(including armchair honeycomb for $t_d=0$) there are no edge states.
However, Kohmoto and Hasegawa \cite{kohmoto-hasegawa} have shown that edge states emerge
in armchair honeycomb provided $t_l\ne t_r$. In the following subsection we reproduce this result
for general anisotropic square lattice.

\subsection{Zero mode edge states}

We discuss zero mode ($E=0$) solutions to (7). The corresponding equations in the component form look as
\babc
\be
(t_u+t_de^{+2ika})\psi_{\bullet n}+t_r\psi_{\bullet n-1}+t_le^{+2ika}\psi_{\bullet n+1}=0\hh,
\ee
\be
(t_u+t_de^{-2ika})\psi_{\circ n}+t_r\psi_{\circ n+1}+t_le^{-2ika}\psi_{\circ n-1}=0\hh,
\ee
\eabc
where $\psi_{\bullet0}=\psi_{\circ0}=0$ and $\psi_{\bullet N+1}=\psi_{\circ N+1}=0$ are assumed.

Solutions to (32) can be written in various forms.
Assuming $t_r\leqslant t_l$ the most appropriate form is (up to normalization)
\babc
\begin{align}
\psi_{\bullet n}&=e^{-in(ka+\pi)}\bigg[\frac{t_r}{t_l}\bigg]^{\frac12n}U_{n-1}\bigg(\frac{t_ue^{-ika}+t_de^{+ika}}{2\sqrt{t_rt_l}}\bigg),\\
\psi_{\circ n}&=e^{-in(ka+\pi)}\bigg[\frac{t_r}{t_l}\bigg]^{\frac12(N-n)}U_{N-n}\bigg(\frac{t_ue^{+ika}+t_de^{-ika}}{2\sqrt{t_rt_l}}\bigg),
\end{align}
\eabc
where the boundary conditions $\psi_{\bullet0}=\psi_{\circ N+1}=0$ are satisfied due to the definition $U_{-1}(x)=0$.
The ones $\psi_{\bullet N+1}=\psi_{\circ0}=0$ lead to a single equation
\be
U_N\bigg(\frac{t_ue^{-ika}+t_de^{+ika}}{2\sqrt{t_rt_l}}\bigg)=0.
\ee

Provided the zeroes of $U_n(x)$ are given by $x_j=\cos\frac{\pi j}{n+1}$ ($j=1,2,\ldots,N$) we resolve (34) as
\be
\frac{t_ue^{-ika}+t_de^{+ika}}{2\sqrt{t_rt_l}}=\cos\frac{\pi j}{N+1}.
\ee
Substituting this into (33) and using (15) we find
\babc
\begin{align}
\psi_{\bullet n}&=(-1)^ne^{-inka}\h\bigg(\frac{t_r}{t_l}\bigg)^{n/2}\h\frac{\sin\frac{\pi nj}{N+1}}{\sin\frac{\pi j}{N+1}}\hh,\\
\nn\\
\psi_{\circ n}&=(-1)^ne^{-inka}\h\bigg(\frac{t_r}{t_l}\bigg)^{(N-n)/2}\h\frac{\sin\frac{\pi nj}{N+1}}{\sin\frac{\pi j}{N+1}}\hh.
\end{align}
\eabc
where irrelevant multiplicative factor is omitted in (36b).

Provided $t_r<t_l$ the wave function $\psi_{\bullet n}$ is exponentially suppressed from the left edge towards
the bulk due to the factor of $(t_r/t_l)^{n/2}$. Analogously, $\psi_{\circ n}$ is suppressed from the right edge towards the bulk.
For $t_r>t_l$ the function $\psi_{\bullet n}$ is localized at the right edge, while $\psi_{\circ n}$ at the left edge.
For $t_r=t_l$ suppression disappears so the edge states never occur.

Due to the trigonometric factors the moduli of these wave functions oscillate with respect to $n$ as shown in Fig. 5.
Note that such oscillations are absent in the edge states observed in zigzag honeycomb.
\begin{figure}[h]
\begin{center}
\includegraphics{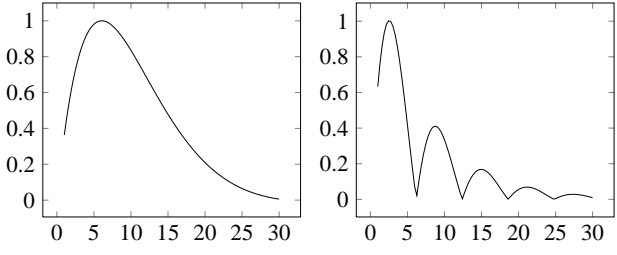}
\end{center}
\caption{$|\psi_{\bullet n}|$ versus $n$ for $N=30$ and $t_r=0.9t_l$ with $j=1$ (left) and $j=5$ (right).}
\end{figure}

We end this subsection by discussing the condition (35) required the zero modes (36) would occur at all.
Apparently the left hand side of (35) must be real, hence there are two cases.

$\bullet$ $k=0$. In this case we find
\be
\frac{t_u+t_d}{2\sqrt{t_rt_l}}=\cos\frac{\pi j}{N+1}.
\ee

$\bullet$ $t_u=t_d$. In this case we come to
\be
\cos(ka)=\frac{\sqrt{t_rt_l}}{t_u}\cos\frac{\pi j}{N+1}.
\ee

We comment on the first case which for $t_d=0$ turns into a honeycomb with armchair edges
($t_d=0$ is unacceptable in the second case where $t_u=t_d$).

Remark, that (37) imposes the following restriction
\be
t_u+t_d\leqslant2\sqrt{t_rt_l}.
\ee

Summarizing, the condition (37) and hence (39) are necessary for occurrence of the zero mode, while $t_r\ne t_l$ is necessary
this zero mode would be localized at the edges.

\section{Anisotropic triangular ribbon}

We consider triangular anisotropic ribbons with three different types of boundaries:
1) linear, 2) single side zigzag and 3) two side zigzag cases as shown in Fig. 6.
These are all soluble in terms of Chebyshev polynomials. We consider them separately in the following subsections.

\begin{figure}[h]
\begin{center}
\includegraphics{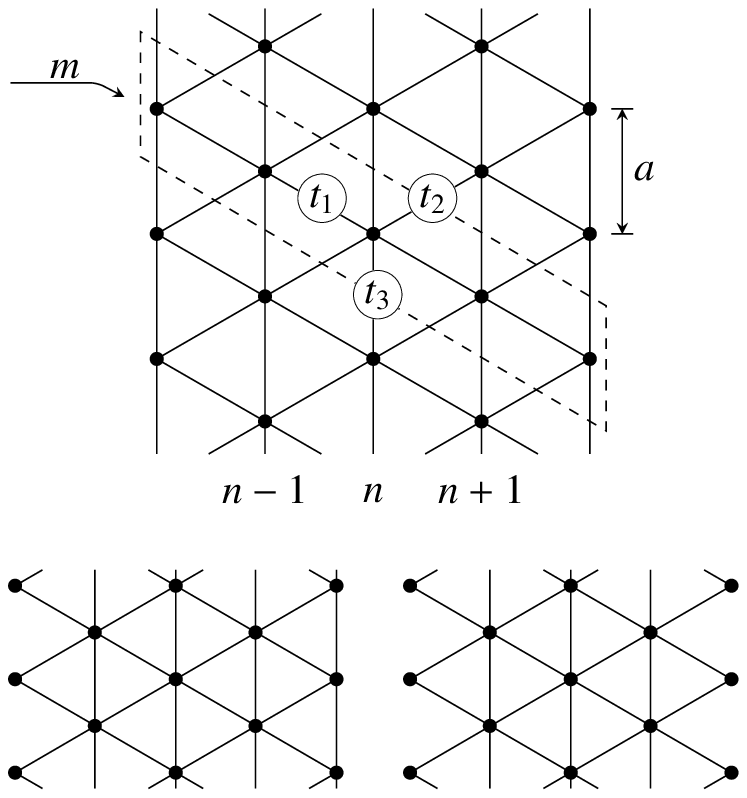}
\end{center}
\caption{Triangular-lattice ribbons with different boundaries: linear edges (upper),
single side zigzag (lower left) and two side zigzag (lower right). All three cases are periodic in $y$-direction with periodicity $a$.}
\end{figure}

\subsection{Linear edges}

In this case (upper panel Fig. 6) the tight-binding Hamiltonian is given by
\begin{align}
H&=t_1\sum_m\sum_{n=2}^{N}\Big[c^\dag(n-1,m)c(n,m)+h.c.\Big]+\nn\\
&+t_2\sum_m\sum_{n=1}^{N-1}\Big[c^\dag(n+1,m+1)c(n,m)+h.c.\Big]+\nn\\
&+t_3\sum_m\sum_{n=1}^{N}\Big[c^\dag(n,m-1)c(n,m)+h.c.\Big].
\end{align}
Employ the Fourier transform
\be
c(n,m)=\frac{1}{\sqrt{2\pi/a}}\int_{BZ} e^{+ikam}c_n(k)dk
\ee
where the width of Brillouin zone is $2\pi/a$.

Then the Hamiltonian (40) takes the form
\be
H=\int\psi^\dag(k)\hh\mathscr H(k)\hh\psi(k)dk
\ee
where $\psi=(c_1,\ldots,c_N)$ and
\be
\mathscr H=2t_3\cos(ka)+\zeta^*\beta+\zeta\beta^\dag
\ee
with $\zeta=t_1+t_2e^{-ika}$ and $\beta$ given by (6).

The eigenvalue equation takes the form
\be
\left(\ba{cccccc}
w&-\zeta^*&\g{0}&\cdots&\g{0}&\g{0}
\\
-\zeta&w&-\zeta^*&\cdots&\g{0}&\g{0}
\\
\g{0}&-\zeta&w&\cdots&\g{0}&\g{0}
\\
\vdots&\vdots&\vdots& &\vdots&\vdots
\\
\g{0}&\g{0}&\g{0}&\cdots&w&-\zeta^*
\\
\g{0}&\g{0}&\g{0}&\cdots&-\zeta&w
\ea\right)\psi=0
\ee
where $w=E-2\hh t_3\cos(ka)$.

Eigenvalues and eigenstates are given by
\babc
\begin{align}
&E_j=2\hh t_3\cos(ka)+2|\zeta|\cos\frac{\pi j}{N+1}\\
\nn\\
&(\psi_j)_n=e^{+i(n-1)\vartheta}\frac{\sin\frac{\pi jn}{N+1}}{\sin\frac{\pi j}{N+1}}\h\psi_1
\end{align}
\eabc
where $\vartheta=arg(\zeta)$ and $j=1,\ldots,N$ labels the eigenstates.

Form (45b) it is obvious that eigenstates exhibit oscillations with respect to $n$, i.e. these are bulk states.

\subsection{Single side zigzag}

We consider the case shown in the lower left panel of Fig. 6.
The corresponding Hamiltonian is obtained by removing the $n=1$ term from the $t_3$-piece of (40).
The eigenvalue equation takes the form
\be
\left(\ba{cccccc}
w+\tau&-\zeta^*&\g{0}&\cdots&\g{0}&\g{0}
\\
-\zeta&w&-\zeta^*&\cdots&\g{0}&\g{0}
\\
\g{0}&-\zeta&w&\cdots&\g{0}&\g{0}
\\
\vdots&\vdots&\vdots& &\vdots&\vdots
\\
\g{0}&\g{0}&\g{0}&\cdots&w&-\zeta^*
\\
\g{0}&\g{0}&\g{0}&\cdots&-\zeta&w
\ea\right)\psi=0
\ee
where $w=E-2\hh t_3\cos(ka)$ and $\tau=2\hh t_3\cos(ka)$.

Secular equation appears as
\be
U_N\bigg(\frac{E-\tau}{2|\zeta|}\bigg)+\frac{\tau}{|\zeta|}U_{N-1}\bigg(\frac{E-b}{2|\zeta|}\bigg)=0
\ee
and determines the eigenvalues $E_1,\ldots,E_N$.
The corresponding eigenstates (up to normalization) are
\be
(\psi_j)_n=e^{+in\vartheta}\bigg[U_{n-1}\bigg(\frac{E_j-\tau}{2|\zeta|}\bigg)+\frac{\tau}{|\zeta|}U_{n-2}\bigg(\frac{E_j-\tau}{2|\zeta|}\bigg)\bigg].
\ee

We are mainly interested in revealing the conditions necessary for the formation of edge states.
Reminding the relation (15) we conclude that for $-1<\frac{E-\tau}{2|\zeta|}<1$
the eigenstates (48) oscillate with respect to $n$ and therefore represents bulk states. Consequently,
the edge states may occur only in the following two cases
\be
\frac{E-\tau}{2|\zeta|}=\pm\cosh u.
\ee
We examine if these conditions can be satisfied by the energy bands determined by (47).

Substituting (49) into (47) and using (20) we find
\be
\frac{\tau}{|\zeta|}=\mp\frac{\sinh[(N+1)u]}{\sinh[Nu]}.
\ee

Squaring up this relation and using the explicit expressions
$\tau=2t_3\cos(ka)$ and $|\zeta|^2=t_1^2+t_2^2+2t_1^{}t_2^{}\cos(ka)$
we arrive to quadratic equation with respect to $\cos(ka)$. Two solutions
corresponding to "$\pm$" signs in (49) are
\be
\cos(ka)=\frac{t_1^{}t_2^{}\mp\sqrt{t_1^2t_2^2+A^2(t_1^2+t_2^2)t_3^2}}{A^2t_3^2}
\ee
\be
A=\frac{2\sinh[Nu]}{\sinh[(N+1)u]}.
\ee
Without loss of generality we assume $t_{1,2,3}>0$, so the upper and lower signs in (51) correspond to $\tau<0$ and $\tau>0$ in (50).

The formal solutions (51) make sense only if the right hand sides are in the interval $[-1,+1]$.
This requirement leads to
\be
\bigg|\frac{t_1\mp t_2}{2t_3}\bigg|<\frac{\sinh[Nu]}{\sinh[(N+1)u]}
\ee
which can be realized for certain values of $u$ only if the following conditions are satisfied
\be
\bigg|\frac{t_1\mp t_2}{2t_3}\bigg|<\frac{N}{N+1}.
\ee

Provided (54) is held, the edge states are parameterized by the values of $u$ satisfying (53).
The corresponding momentum $k$ and energy $E$ are determined by (51) and (49).
Eigenstates can be obtained by substituting (49) into (48) and using (20).
These appear as
\be
\psi_n=(\pm1)^{n-1}e^{+in\vartheta}\frac{\sinh[(N-n+1)u]}{\sinh[Nu]}.
\ee

Fig. 7 shows $E$ versus $k$ for $t_{1,2,3}=0.9,0.1,1$ with $N=5$.
\begin{figure}
\begin{center}
\includegraphics{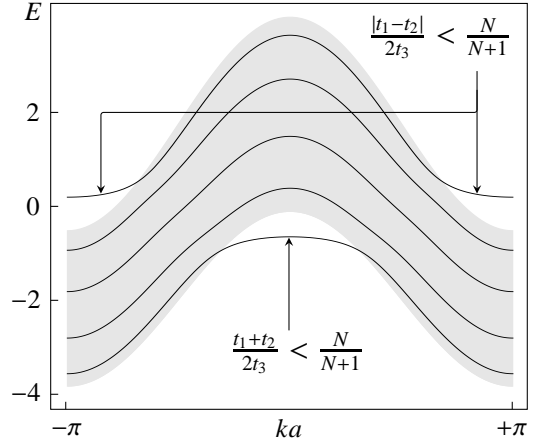}
\end{center}
\caption{Energy spectrum for $N=5$ and $t_{1,2,3}=0.9,0.1,1$. Shaded area is bounded in vertical directions by
$\frac{E-\tau}{2|\zeta|}=\pm1$, so that in the interior we have $-1<\frac{E-\tau}{2|\zeta|}<1$.
Therefore the energy band segments located within the shaded area are bulk states, while the ones beyond represent the edge states.
For the particular values of hopping parameters the relations (54) are both satisfied. Correspondingly, we have two subbands of edge states.
One of them located below the shaded area occurs due to $\frac{t_1+t_2}{2t_3}<\frac{N}{N+1}$, while the other above the area appears
due to $\frac{|t_1-t_2|}{2t_3}<\frac{N}{N+1}$.}
\end{figure}
Wave functions (55) are plotted in Fig. 8 where from it is obvious that localization occurs near $n=1$, i.e. at zigzag edge.
\begin{figure}[h]
\begin{center}
\includegraphics{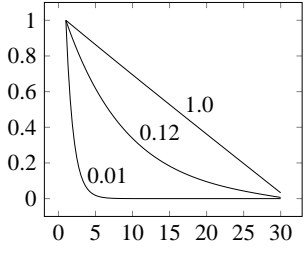}
\end{center}
\caption{Edge wave function (55) versus $n$ for $N=30$ and different values of $u$. Phase factor of $(\pm1)^{n-1}e^{+in\vartheta}$ omitted.}
\end{figure}

\subsection{Two side zigzag}

We next consider the case depicted in lower right panel of Fig. 6.
The corresponding Hamiltonian is obtained by removing the $n=1$ and $n=N$ terms from the $t_3$-piece in (40).
The eigenvalue equation takes the form
\be
\left(\ba{cccccc}
w+\tau&-\zeta^*&\g{0}&\cdots&\g{0}&\g{0}
\\
-\zeta&w&-\zeta^*&\cdots&\g{0}&\g{0}
\\
\g{0}&-\zeta&w&\cdots&\g{0}&\g{0}
\\
\vdots&\vdots&\vdots& &\vdots&\vdots
\\
\g{0}&\g{0}&\g{0}&\cdots&w&-\zeta^*
\\
\g{0}&\g{0}&\g{0}&\cdots&-\zeta&w+\tau
\ea\right)\psi=0
\ee
where  $w=E-\tau$ and $\tau=2\hh t_3\cos(ka)$.

Compared to (46) only the last line is modified. As shown in Appendix the last line determines secular equation while
the rest lines determine the eigenstate components. Therefore the eigenstate expressions are the same as in the case
of single side zigzag
\be
(\psi_j)_n=e^{+in\vartheta}\bigg[U_{n-1}\bigg(\frac{E_j-\tau}{2|\zeta|}\bigg)+\frac{\tau}{|\zeta|}U_{n-2}\bigg(\frac{E_j-\tau}{2|\zeta|}\bigg)\bigg],
\ee
while the secular equation appears as
\begin{align}
U_N\bigg(\frac{E-\tau}{2|\zeta|}\bigg)
&+\frac{2\tau}{|\zeta|}\h U_{N-1}\bigg(\frac{E-\tau}{2|\zeta|}\bigg)+\nn\\
\nn\\
&+\frac{\tau^2}{|\zeta|^2}U_{N-2}\bigg(\frac{E-\tau}{2|\zeta|}\bigg)=0.
\end{align}

Searching for the edge states we employ the same arguments as for single side zigzag edges, i.e. we introduce
\be
\frac{E-\tau}{2|\zeta|}=\pm\cosh u.
\ee
Substituting into (58) we come to
\be
\frac{\sinh[(N+1)u]}{\sinh[(N-1)u]}\pm\frac{2\sinh[Nu]}{\sinh[(N-1)u]}\frac{\tau}{|\zeta|}+\frac{\tau^2}{|\zeta|^2}=0,
\ee
which gives the following four solutions
\babc
\begin{align}
\cos(ka)=\frac{t_1^{}t_2^{}\mp\sqrt{t_1^2t_2^2+A^2(t_1^2+t_2^2)t_3^2}}{A^2t_3^2},\\
\nn\\
\cos(ka)=\frac{t_1^{}t_2^{}\mp\sqrt{t_1^2t_2^2+B^2(t_1^2+t_2^2)t_3^2}}{B^2t_3^2},
\end{align}
\eabc
where
\babc
\begin{align}
A=\frac{2\sinh[(N-1)u]}{\sinh(Nu)-\sinh u},\\
\nn\\
B=\frac{2\sinh[(N-1)u]}{\sinh(Nu)+\sinh u}.
\end{align}
\eabc

Requiring the right hand sides of (61) to lay in the interval $[-1,+1]$ we obtain
\babc
\begin{align}
\bigg|\frac{t_1\mp t_2}{2 t_3}\bigg|<\frac{\sinh[(N-1)u]}{\sinh(Nu)-\sinh u}\\
\nn\\
\bigg|\frac{t_1\mp t_2}{2 t_3}\bigg|<\frac{\sinh[(N-1)u]}{\sinh(Nu)+\sinh u}
\end{align}
\eabc
for (61a) and (61b) respectively.

These can be satisfied for certain values of $u$ only if
\babc
\begin{align}
\bigg|\frac{t_1\mp t_2}{2 t_3}\bigg|&<1\\
\nn\\
\bigg|\frac{t_1\mp t_2}{2 t_3}\bigg|&<\frac{N-1}{N+1}
\end{align}
\eabc
respectively.

Substituting (59) into (57) and using (61) in $\tau$ and $|\zeta|$ yields
\babc
\be
\psi_n=(\pm1)^{n-1}e^{-in\vartheta}\frac{\sinh[(N-n)u]+\sinh[(n-1)u]}{\sinh[(N-1)u]},
\ee
\be
\psi_n=(\pm1)^{n-1}e^{-in\vartheta}\frac{\sinh[(N-n)u]-\sinh[(n-1)u]}{\sinh[(N-1)u]},
\ee
\eabc
for (64a) and (64b) respectively.

Thus, we may have up to four segments of $k$ representing edge states.
Edge states emerge in these intervals of $k$ only if the corresponding condition from (64) is satisfied.
In Fig. 9 we plot $E$ versus $k$ for $N=5$ and various values of $t_1,t_2,t_3$. Wave functions (65) are plotted in Fig. 10.
\begin{figure}[h]
\begin{center}
\includegraphics{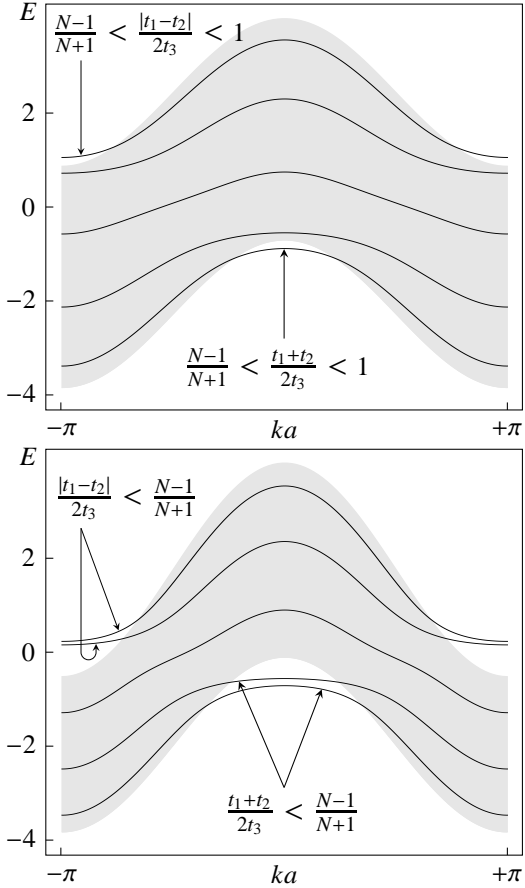}
\end{center}
\caption{Energy $E$ versus $k$ for $N=5$ with $t_{1,2,3}=1.5,0.1,1$ (upper) and $t_{1,2,3}=0.9,0.1,1$ (lower).
Energy band segments laying beyond the shaded area represent the edge states.
Inequalities express the conditions when the corresponding segments appear.}
\end{figure}
\begin{figure}[h]
\begin{center}
\includegraphics{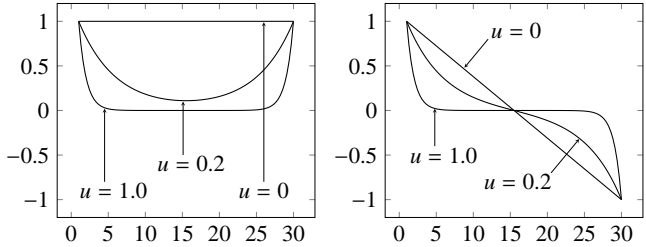}
\end{center}
\caption{Edge state wave functions (65a) (left) and (65b) (right) versus $n$ for $N=30$ and different values of $u$.
Phase factors of $(\pm1)^{n-1}e^{-in\vartheta}$ are omitted.}
\end{figure}

\section{Conclusions}

In this paper we have considered tight-binding models on particular class of lattice ribbons where
the eigenvalue problems lead to three-term recurrence relations. Such a selection is motivated
by the fact that three-term recurrence relations are usually resolved by orthogonal polynomials,
which in the cases under consideration turn to be the Chebyshev polynomials of the second kind.
The technique developed is capable of handling ribbons with hopping anisotropy. Within the given
approach we have reproduced the results due to Wakabayashi et al. \cite{Kobayashi_et_al_Review_2010}
for isotropic honeycomb ribbons with zigzag and armchair edges, and the one due to Kohmoto and
Hasegawa \cite{kohmoto-hasegawa} for zero mode edge states in anisotropic armchair honeycomb.
Inclusion of hopping anisotropy allowed to trace out the corresponding influence on the formation of edge states.
Also, anisotropic triangular ribbons with various edge geometries are studied within the same approach.

\section*{Acknowledgements}

We are grateful to D. Baeriswyl and M. Sekania for illuminating discussions.

\appendix

\section{}

We comment on solving (10a). We first get rid of the phases of $\xi$ by taking $\psi_{\circ n} =e^{-in\vartheta}\phi_n$
with $\vartheta=arg(\xi)$. Then the equation written out in components appears as
\begin{align}
|\xi|\phi_2=&\tilde w\phi_1\nn\\
|\xi|\phi_3=&w\phi_2-|\xi|\phi_1\nn\\
|\xi|\phi_4=&w\phi_3-|\xi|\phi_2\nn\\
&\vdots\\
|\xi|\phi_N=&w\phi_{N-1}-|\xi|\phi_{N-2}\nn\\
0=&w\phi_{N}-|\xi|\phi_{N-1}\nn
\end{align}

The system is homogeneous hence comprises one undetermined constant we choose to be $\phi_1$.
Then $\phi_2$ can be solved out from the first equation. Substituting this into the second we solve out $\phi_3$
and so on. Using induction method we can show the following relation
\be
\phi_{n+1}=D_n\phi_1
\ee
where
\be
D_n\equiv det\left(\ba{cccccc}
\displaystyle\frac{\tilde w}{|\xi|}&1&\g{0}&\cdots&\g{0}&\g{0}
\\
1&\displaystyle\frac{w}{|\xi|}&1&\cdots&\g{0}&\g{0}
\\
\g{0}&1&\displaystyle\frac{w}{|\xi|}&\cdots&\g{0}&\g{0}
\\
\vdots&\vdots&\vdots& &\vdots&\vdots
\\
\g{0}&\g{0}&\g{0}&\cdots&\displaystyle\frac{w}{|\xi|}&1
\\
\g{0}&\g{0}&\g{0}&\cdots&1&\displaystyle\frac{w}{|\xi|}
\ea\right)_{n\times n}
\ee

Expression (A2) with $n=1,\ldots,N-1$ resolves the first $N-1$ equations of (A1),
while the last equation implies
\be
D_N=0
\ee
and gives the eigenvalues $\omega_1,\ldots,\omega_N$.

We now calculate $D_n$. We first calculate
\be
A_n\equiv det\left(\ba{cccccc}
\displaystyle\frac{w}{|\xi|}&1&\g{0}&\cdots&\g{0}&\g{0}
\\
1&\displaystyle\frac{w}{|\xi|}&1&\cdots&\g{0}&\g{0}
\\
\g{0}&1&\displaystyle\frac{w}{|\xi|}&\cdots&\g{0}&\g{0}
\\
\vdots&\vdots&\vdots& &\vdots&\vdots
\\
\g{0}&\g{0}&\g{0}&\cdots&\displaystyle\frac{w}{|\xi|}&1
\\
\g{0}&\g{0}&\g{0}&\cdots&1&\displaystyle\frac{w}{|\xi|}
\ea\right)_{n\times n}
\ee
which differs from $D_n$ by $w$ instead of $\tilde w$ in the upper left corner.
Expanding (A5) with respect to first row we find
\be
A_n=\frac{w}{|\xi|}\hh A_{n-1}-A_{n-2}.
\ee

Remind that the Chebyshev polynomials $U_n(x)$ satisfy the recurrence relations
$U_n(x)=2xU_{n-1}(x)-U_{n-2}(x)$. Comparing this with (A6) we come to
\be
A_n=U_n\bigg(\frac{w}{2|\xi|}\bigg).
\ee

Expanding (A3) with respect to the first row we obtain
\be
D_n=\frac{\tilde w}{|\xi|}\hh A_{n-1}-A_{n-2}.
\ee
Using $\tilde w=w+1$ together with (A6) and (A7) we find
\be
D_n=U_n\bigg(\frac{w}{2|\xi|}\bigg)+\frac{1}{|\xi|}\hh U_{n-1}\bigg(\frac{w}{2|\xi|}\bigg).
\ee
Combining this with (A2) and $\psi_{\circ n}=e^{-in\vartheta}\phi_n^{}$ we obtain (12a).
The secular equation (A4) takes the form (11).

Equation (10b) is solved in the same way by expressing all components via $\psi_{\bullet N}$
(starting from the lower right corner instead of the upper left).

\end{document}